\documentstyle[a]{article}

\input{psfig.sty}

\def\N{N_c}
\def\m{m_Q}
\def\L{\Lambda_{QCD}}

\def\l{\lambda}
\def\c{\cite}

\begin{document}

\title{HEAVY BARYONS: \\A COMBINED LARGE $\N$ AND HEAVY QUARK \\
EXPANSION FOR ELECTROWEAK CURRENTS}

\author{Boris A. Gelman}

\address{Department of Physics, University of Maryland, College Park\\
College Park, MD 20742-4111, USA\\E-mail: gelmanb@physics.umd.edu}


\maketitle

\abstracts{The combined large $\N$ and heavy quark limit for
baryons containing a single heavy quark is discussed. The
combined large $\N$ and heavy quark expansion of the heavy quark
bilinear operators is obtained. In the combined expansion the
corrections proportional to $m_N/\m$ are summed to all orders. In
particular, the combined expansion can be used to determine
semileptonic form factors of heavy baryons in the combined limit.}

\section{Introduction}
The hadrons are strongly bound states of quarks and gluons. Their
description directly from QCD---the underlying theory of strong
interactions---is a daunting task significantly impeded by the
highly non-perturbative nature of QCD at low energies. Lattice QCD
is the only fully non-perturbative framework at the present time
to calculate the low energy hadronic observables from QCD. Its
present capabilities, however, are at a rudimentary level to
perform fully dynamical (non-quenched) calculations of
interesting QCD and electroweak observables. In the absence of
the direct analytical methods, we turn to the approximations which
lead to model independent predictions.

A popular approach to describe systems at low-energies in a model
independent way is to use the methods of effective field theories
(EFT) \c{EFT}. These methods are most successful when the spectrum
of the underlying theory exhibits a large enough scale separation
so that the low-energy interactions of relevant degrees of freedom
can be described in terms of a small number of operators with the
remainder being suppressed by higher powers of the ratio
$E/\Lambda$, where $E$ and $\Lambda$ are typical low-energy and
high-energy scales. A power counting scheme---a critical
ingredient of EFT---selects all dominant interactions at a given
order in the expansion in powers of $E/\Lambda$. A finite number
of undermined constants at each order can be found from
experiment and used to predict the outcome of others.

The symmetries of the system of interest greatly constrain
model-independent power counting schemes. It is often the case
that such symmetries are approximate and manifest themselves in a
particular limit of QCD. An example is the well-known chiral
symmetry in the light-quark sector of QCD which appears in the
zero quark mass limit. Chiral perturbation theory---an effective
theory describing interactions of pions and nucleons---is based on
power counting emerging from the underlying chiral symmetry.

Another example is heavy quark effective theory (HQET) \c{HQET}
used to describe hadrons containing a single heavy quark---the
{\it heavy mesons} and {\it heavy baryons}. HQET is based on the
heavy quark spin-flavor symmetry which emerges in the heavy quark
limit, {\it i.e.} when the heavy quark mass, $\m$, goes to
infinity. The heavy quark effective Lagrangian has a form of an
expansion in powers of $k/\m$, where $k$ (${\cal O}(\L)$) is a
typical momentum of the light degrees of freedom inside the heavy
hadron. To stress the non-perturbative nature of these degrees of
freedom they are often collectively referred to as the {\it brown
muck}.

The heavy quark spin-flavor symmetry relates the heavy hadrons
moving at the same 4-velocity. The symmetry is described by the
unitary transformations corresponding to $SU(2\,n_f)$ group,
where $n_f$ is the number of heavy flavors. This symmetry implies
a number of important phenomenological consequences. In the heavy
baryon sector of QCD, which is of the primary interest here, the
heavy quark symmetry restricted to charm and bottom flavors
predicts, for example, an existence of the degenerate heavy quark
spin doublets. An approximately degenerate doublet of the excited
charm baryons has been observed---($\Lambda_{c1},\,
\Lambda^{*}_{c1}$). The symmetry breaking is of order of $1/\m$
as expected. Heavy quark symmetry also leads to the relation
between electroweak matrix elements that determine the
semileptonic decay form factors of heavy baryons. In particular,
the normalization of these form-factors at zero recoil is given
by the conserving heavy quark number \c{HBHQET}.

Here we will discuss a combined heavy quark and large $\N$ limit
for the heavy baryon sector of QCD. It turns out, in this limit
the spectrum of heavy baryons exhibits an approximate symmetry
associated with a contracted $O(8)$ group \c{hb0,hb1}. This
symmetry connects the low-energy excited states of heavy baryons
to the ground state. An effective field theory can be derived to
describe these low-energy excited states and their electroweak
decays \c{hb2}.

Large $\N$ QCD \c{LN1} is a very useful framework to obtain model
independent predictions for baryons \c{LN2}. In particular, the
baryons containing only $u$ and $d$ quarks in the large $\N$
exhibit light quark spin-flavor symmetry described by the
contracted $SU(4)$ group \c{SF}. An infinite dimensional
irreducible representation of this symmetry is given by baryons
with the spin-isospin quantum numbers $I=J=1/2, 3/2, 5/2$, etc.
The two lowest states $I=J=1/2, 3/2$ correspond to the large $\N$
analogs of the nucleon and $\Delta$. The baryons in the large
$\N$ limit arise as solitons of the non-linear chiral Lagrangian
\c{LN2}.

\section{The Bound State Picture}
The attractiveness of the combined heavy quark and large $\N$
limit can be seen in a simple model, or more precisely, the type
of models that have been considered in the past \c{bst}. In these
models the heavy baryon is thought of as the bound state of a
heavy meson and an ordinary baryon. For example, the $\Lambda_c$
baryon is thought of as bound state of $D$ or $D^*$ meson and a
nucleon. As both $\m$ and $\N$ tend to infinity, the difference
between the heavy quark mass, $\m$, and the heavy meson mass,
$m_H$, as well as the difference between the brown muck mass and
the nucleon mass, $m_N$, are suppressed by powers of $1/\N$ and
$1/\m$.

The combined limit is taken in such a way as to hold the ratio
$\N \Lambda/\m$ constant but arbitrary (here $\Lambda$ represents
a typical hadronic scale of order of $1\, GeV$). In other words,
the results are independent of the order in which two limits are
taken. The corrections to the limiting values are given in powers
of small dimensionless parameter $\lambda$,
\begin{equation}
\l \sim {\Lambda \over \m}\, , {1\over \N} \label{l} \, .
\end{equation}

In the combined heavy quark and large $\N$ limit both $\m$ and
$m_N$ go to infinity. On the other hand, as shown by Witten
\c{LN2}, the meson-baryon interaction is of order of $N_{c}^{0}$.
As a result, the two heavy particles, the heavy meson and a
nucleon, are bound by the potential of order unity. The lowest
excitations of such a system correspond to excitations of the
three dimensional harmonic oscillator with the {\it spring
constant} $\kappa$ of order $\l^0$ and mass $\mu=m_N \m
/(m_N+\m)\sim \l^{-1}$. The energy splitting between the
low-lying excited states is $(\kappa/\mu)^{1/2}$ which is of
order $\l^{1/2} \L$. Hence, as $\l$ goes to infinity in the
combined limit, the low-lying states become degenerate and
exhibit a new symmetry. This symmetry is described by a
contracted $O(8)$ group.

This symmetry has been shown to arise in a model-independent way
in the heavy baryon sector in the combined limit \c{hb0,hb1}. This
symmetry can be used to construct an effective theory in the
combined limit which describes low-energy excited states of heavy
baryons \c{hb2}. The dominant excitations in the combined limit
are the harmonic modes of the collective motion of the brown muck
as whole relative to the heavy quark. These excitations are of
order $\l^{1/2} \L$ while the excitations of the brown muck ({\it
e.g.} the rotation in the isospin space) are of order $\l^0 \L$.

The effective Hamiltonian can be derived from QCD using the
operators which excite the collective motions of the brown muck
relative to the heavy quark. It turns out that the effective
Hamiltonian has the form of an expansion in powers of $\l^{1/2}$.
The leading non-trivial terms exhibit a contracted $O(8)$
symmetry and contain one undetermined parameter corresponding to
a {\it spring constant} $\kappa$. At next-to-leading order an
additional parameter must be fit to experiment. Additional
observables can be obtained from the effective expansion of the
heavy baryon electroweak currents. This does not introduce any
new constants at leading and next-to-leading order in $\l^{1/2}$.

The effective theory can be used to determine a number of heavy
baryon observables, namely masses of the excited states,
semileptonic form factors and electromagnetic decay rates. At
leading order these predictions coincide with corresponding ones
obtained from bound state models. The corrections at
next-to-leading order are expected to be ${\cal O}(1/\N)$.

Here we will describe a method to obtain a combined heavy quark
and large $\N$ expansion of the heavy baryon electroweak current
needed to calculate semileptonic form factors \c{hb2}.
\section{A Combined Expansion of the Heavy Quark Currents}
In HQET the $1/\m$ corrections include those that are
proportional to $m_N/\m$. These corrections are suppressed in
HQET. However, in the combined limit $m_N$ is formally of the
same order as $\m$, so that the convergence of the heavy quark
expansion is spoiled. Note, that the ratio $m_N/\m$ is
approximately $1/2$ for charmed meson, which is not much smaller
unity. This may have some phenomenological consequences. In fact,
the effective theory in the combined limit predicts the ratio of
the charm to bottom electromagnetic decay rates to be
approximately $0.2$ at leading order while the leading order HQET
value is unity \c{hb3,modind}. It is desirable, therefore, to
improve the convergence of the effective expansion in the
combined limit.

The effective $1/\m$ expansion in HQET is obtained by integrating
out the types of the heavy quark interactions that are $1/\m$
suppressed relative to the leading terms. One effect corresponds
to the heavy quark spin interactions with the chromomagnetic
field. The suppression is made explicit by splitting the heavy
quark field in two parts using projection operators
$P_{\pm}=(1\pm \!\!\not\!\! v)/2$. The lower components of the
heavy quark field are $1/\m$ suppressed. In the combined heavy
quark and large $\N$ limit such suppression holds as well.

Another suppressed effect comes from the off-shell momentum of the
heavy quark due to its interaction with the brown muck, $k=p-\m v
\sim \L$, where $p$ and $v$ are the heavy quark 4-momentum and
4-velocity. In the heavy quark limit $p$ and $v$ are almost equal
to the total 4-momentum and 4-velocity of the heavy hadron. In
the combined limit, however, the brown muck contribution to the
total 4-momentum is formally of the same order as that of the
heavy quark since $m_N \sim \m \sim \l^{-1}$. Thus, it is more
appropriate to define the off-shell momentum of the heavy quark
as $k=P-(\m +m_N)v \sim \L$, where $P$ is the total 4-momentum of
the heavy hadron.

The suppression of the heavy quark off-shell fluctuations is made
explicit in HQET by redefining a phase of the heavy quark field:
\begin{equation}
h_Q (x)= e^{i\m v\cdot x} P_{+} Q(x) \,,\,\,\,\,\,\, H_Q (x)=
e^{i\m v\cdot x} P_{-} Q(x) \, \label{hH}
\end{equation}
where $Q(x)$ is the total heavy quark field and $v$ is the
4-velocity of the heavy hadron. The operators $P_{\pm}=(1\pm
\!\!\not\!\! v)/2$ project the total heavy quark field in two
parts. The $H_Q (x)$ component is suppressed by $1/\m$ relative
to $h_Q (x)$ which can be seen from equations of motions written
in terms of these fields.

The typical 4-momentum carried by the field $h_Q (x)$ in the
heavy hadron state is $k \sim M-\m$, where $M$ is the total mass
of the hadron, so that in HQET $k$ is ${\cal O}(\L/\m)$. However,
in the combined limit $k\sim m_N\sim \l^{-1}$ and is not
suppressed. In order to remove the brown muck contribution to the
off-shell momentum one can redefine the effective fields $h_Q
(x)$ and $H_Q (x)$ by changing their phases:
\begin{eqnarray}
h_Q (x)& \rightarrow & h^{\l}_Q (x)= e^{i(\m+m_N) v\cdot x} P_{+}
Q(x)
\, , \nonumber \\
H_Q (x) & \rightarrow & H^{\l}_Q (x)= e^{i(\m+m_N) v\cdot x} P_{-}
Q(x) \, \label{hHcombined} \, ,
\end{eqnarray}
where a superscript $\l$ indicates that effective fields
$h^{\l}_Q (x)$ and $H^{\l}_Q (x)$ are used in the combined
expansion. The field $h^{\l}_Q (x)$ defined in
Eq.~(\ref{hHcombined}) carries a typical off-shell 4-momentum $k
\sim M-(\m+m_N)\sim \l^{0}\,\L$ which is suppressed in the
combined limit.

The phase redefinition in Eq.~(\ref{hHcombined}) limits the
off-shell fluctuations of the effective heavy quark fields
$h^{\l}_Q (x)$ and $H^{\l}_Q (x)$. However, it can potentially
spoil relative scaling of the fields $h^{\l}_Q (x)$ and $H^{\l}_Q
(x)$. Indeed, using the equations of motion that follow from the
heavy quark Lagrangian, ${\cal L}_{Q} = \bar{Q}(x)\left(i
\!\not\!\!D-\m\right)Q(x)$, one sees that $H_Q (x) \sim (1/\m)\,
h_Q (x)$ while $H^{\l}_Q (x) \sim (m_N/\m)\, h^{\l}_Q (x)$. Thus,
it seems that new definition, Eq.~(\ref{hHcombined}), prevents the
consistent elimination of $H^{\l}_Q (x)$ in terms of $h^{\l}_Q
(x)$. This seeming inconsistency is resolved, however, by noting
that in the combined limit the equations of motions for $h^{\l}_Q
(x)$ and $H^{\l}_Q (x)$ should be obtained not from ${\cal
L}_{Q}$ but from an effective Lagrangian that takes into account
the large contribution of the brown muck.

What is this Lagrangian? The total Lagrangian has the form:
\begin{equation}
{\cal L}={\cal L}_{Q}+{\cal L}_{q}+{\cal L}_{YM} \,, \label{L}
\end{equation}
where ${\cal L}_{q}$ is the light quark Lagrangian and ${\cal
L}_{YM}$ is the Yang-Mills Lagrangian for gluon fields. In the
combined limit ${\cal L}_{q} +{\cal L}_{YM}$ contains a large
piece corresponding to the brown muck mass $m_N$. This
contribution can be made explicit by adding and subtracting from
the total Lagrangian an effective term $m_N \bar{Q}\!\!\not\!v Q$,
which contains only the heavy quark fields:
\begin{equation}
{\cal L}=({\cal L}_{Q}-m_N \bar{Q}\!\!\not\!v Q)+({\cal
L}_{q}+{\cal L}_{YM}+m_N \bar{Q}\!\!\not\!v Q) \,. \label{Ltotal}
\end{equation}
The operator $\bar{Q}\!\!\not\!v Q$ in the heavy baryon reference
frame counts the number of heavy quarks.

To obtain the correct scaling of $h^{\l}_Q (x)$ and $H^{\l}_Q (x)$
one needs to use equations of motion obtained from an effective
Largangian, ${\cal L}_{eff}={\cal L}_{Q}-m_N \bar{Q}\!\!\not\!v
Q$. Indeed, it can be shown that in this case $H^{\l}_Q$ is ${\cal
O}(\l \, h^{\l}_Q)$ in the combined limit. As a result, $H^{\l}_Q
(x)$ can be consistently eliminated from the effective
Lagrangian, ${\cal L}_{eff}$, and heavy quark operators at each
order in the combined expansion. In fact, the resulting expansion
of the heavy quark current has the same form as the corresponding
expansion in HQET:
\begin{equation}
\bar{c}\, \Gamma \, b = \bar{h}^{\l}_c\, \Gamma\, h^{\l}_b
+{1\over 2m_c}\,\bar{h}^{\l}_c\, \Gamma\, (i\!\not\!\!
D)\,h^{\l}_b -{1\over 2m_b}\,\bar{h}^{\l}_c\, \Gamma\,
(i\overleftarrow{\!\not\!\! D})\,h^{\l}_b + {\cal O}(\l^2) \, ,
\label{current}
\end{equation}
where $\Gamma$ gives a Dirac structure of the current.

The first term in Eq.~(\ref{current}) contributes ${\cal
O}(\l^0)$ to the heavy baryon matrix elements while the two other
terms give ${\cal O}(\l)$ contributions. This happens because the
covariant derivative brings down one power of $k$ which is ${\cal
O}(\l^0)$ if the field $h^{\l}_Q (x)$ is defined as in
Eq.~(\ref{hHcombined}). In other words, the corrections
proportional to $m_N/\m$ are summed to all orders.

The expansion in Eq.~(\ref{current}) can be used to determine the
semileptonic form factors of the heavy baryons in the combined
limit. The effective Hamiltonian in the combined limit has a form
of expansion in powers of $\l^{1/2}$. Hence, at leading and
next-to-leading order only the first term in Eq.~(\ref{current})
contributes to the electroweak matrix elements. The matrix
elements of this leading operator can be determined in the
combined limit. They give the leading contributions to the
semileptonic form factors $\Theta$ and $\Xi$ defined by
\c{hb3,modind}:
\begin{eqnarray}
<\Lambda_c(\vec{v}^{\,\prime})\,|\,\bar{c}\,
\gamma^\mu(1-\gamma_5) \, b\,|\,\Lambda_b(\vec{v})> & = &
\Theta\,\bar{u}_c\, \gamma^\mu(1-\gamma_5) \,
u_b\,(1+{\cal O}(\l^{3/2})) \, , \\
<\Lambda_{c1}(\vec{v}^{\,\prime})\,|\,\bar{c}\,
\gamma^\mu(1-\gamma_5) \, b\,|\,\Lambda_b (\vec{v})> & = &
\sqrt{3}\,\,\Xi\,\,\bar{u}_c\, \gamma^\mu(1-\gamma_5) \,
u_b\,(1+{\cal O}(\l^{3/2})) \, , \nonumber
\\<\Lambda_{c1}^{*}(\vec{v}^{\,\prime})\,|\,\bar{c}\,
\gamma^\mu(1-\gamma_5) \, b\,|\,\Lambda_b (\vec{v})> & = &
\Xi\,\,\bar{u}_{c\nu}\, (\sigma^{\nu\mu}\gamma_5-g^{\mu\nu}) \,
u_b\,(1+{\cal O}(\l^{3/2})) \, , \nonumber \label{FF}
\end{eqnarray}
where $(\Lambda_{c1}, \Lambda_{c1}^{*})$ is a  doublet of the
first excited state of $\Lambda_c$ with $J=1/2$ and $J=3/2$;
$u_{c\nu}$ is a Rarita-Schwinger spinor normalized by
$\bar{u}_{c\nu}\,u_{c}^{\nu}=-1$. Equations~(\ref{FF}) are valid
for the velocity transfers $|\vec{v}^{\,\prime}-\vec{v}|^2$ of
order $\lambda^{3/2}$.
\section{Conclusion}
We have discussed the way to obtain the combined heavy quark and
large $\N$ expansion for the heavy baryon electroweak currents.
The definition of the effective heavy quark fields differs from
the corresponding definition in HQET by a phase factor
(Eq.~(\ref{hHcombined})) proportional to the brown muck mass. The
latter is suppressed in HQET but formally is of the same order as
the heavy quark mass in the combined limit. In the combined
expansion the $m_N/\m$ corrections are summed to all orders.

\section*{Acknowledgments}
This work was supported by the U.S. Department of Energy, under
Grant No. DOE-93ER-40762-251.

\end{document}